# Non-destructive determination of phase, size, and strain of individual grains in polycrystalline photovoltaic materials

Mariana Mar Lucas[1], Tiago Ramos[1], Peter S. Jørgensen[1], Stela Canulescu[3], Peter Kenesei[4], Jonathan Wright[5], Henning F. Poulsen[2], Jens W. Andreasen[1].

[1.] Department of Energy Conversion and Storage, Technical University of Denmark, 2800 Kgs. Lyngby, Denmark.
[2.] Department of Physics, Technical University of Denmark, 2800 Kgs. Lyngby, Denmark.
[3.] Department of Photonics Engineering, Technical University of Denmark, 2800 Kgs. Lyngby, Denmark.
[4.] X-ray Science Division, Argonne National Laboratory, 9700 S Cass Avenue, Lemont IL 60439
[5.] European Synchrotron Radiation Facility, 71 avenue des Martyrs, 38000 Grenoble, France.

## 1  Abstract

We demonstrate a non-destructive approach to provide structural properties on the grain level for the absorber layer of kesterite solar cells. Kesterite solar cells are notoriously difficult to characterize structurally due to the co-existence of several phases with very similar lattice parameters.

Specifically, we present a comprehensive study of 597 grains in the absorber layer of a 1.64% efficient $Cu_2ZnSnS_4$ (CZTS) thin-film solar cell, from which 15 grains correspond to the secondary phase ZnS. By means of three dimensional X-ray diffraction (3DXRD), we obtained statistics for the phase, size, orientation, and strain tensors of the grains, as well as their twin relations. We observe an average tensile stress in the plane of the film of ~ 70 MPa and a compressive stress along the normal to the film of ~ 145 MPa. At the grain level, we derive a 3D stress tensor that deviates from the biaxial model usually assumed for thin films. 41% of the grains are twins. We calculate the frequency of the six types of Σ3 boundaries, revealing that 180° rotations along axis <221> is the most frequent. This technique can be applied to polycrystalline thin film solar cells in general, where strain can influence the bandgap of the absorber layer material, and twin boundaries play a role in the charge transport mechanisms.

## 2  Introduction

Photovoltaic thin-film technology is increasingly targeting alternative materials to meet the triple challenge of sustainability, low energy payback time, and scalability. Current technologies include polycrystalline CdTe [1] and $Cu(In, Ga)Se_2$ (CIGS) [2], both with power conversion efficiencies that surpass 20%. A relatively new but promising candidate is $Cu_2ZnSnS_4$ (CZTS), with an efficiency of 11% [3], and the selenized version, $Cu_2ZnSn(S, Se)_4$, where efficiency has reached 12.6% [4]. All of these materials still perform below the Shockley–Queisser limit [5].

The performance of these materials is strongly dependent on their complex microstructures. One limiting factor shared among these semiconductors is the deficient open-circuit voltage ($V_{OC}$) attributed, among other reasons, to the structural defects in the absorber layer. For example, a small grain size is associated with an increased amount of grain boundaries, which, if poorly passivated, can contribute to carrier recombination [6]–[9]. Secondary phases can cause other deficiencies. For CZTS with a typically Cu-poor and Zn-rich composition, secondary phases with different bandgaps form, such as the high bandgap ZnS, increasing series resistance when situated in the back contact of the solar cell or acting as a barrier to the charge carriers at the p-n junction [10]– [12]. In CIGS absorbers, which usually have a Cu-poor composition, a Cu(In, Ga)$_3$Se$_5$ phase can occur at the surface with a high density of indium or gallium appearing as copper antisite (In, Ga)$_{Cu}$ defects and acting as recombination centers [13]. A lattice mismatch between CIGS and Cu(In, Ga)$_3$Se$_5$ can also create structural defects and an increased density of recombination centers [14]. Moreover, the lattice spacing changes when modifying the material composition while tuning the band gap e.g., with the variation of Ga/In and Se/S ratios.

Furthermore, the structure of the grains and the local strain change inevitably, as the multicomponent materials undergo different treatments from the deposition and the annealing of the absorber layer to the post-treatment methods for the coating of the subsequent layers of the device. The change in lattice parameters, as a result of fabrication stresses, can affect not only the mechanical properties of the film (adhesion to the substrate, elastic modulus, and deformation [15]) but the electronic properties as well. As an example, theoretical calculations demonstrate the reduction of bandgap due to a tensile biaxial in-plane strain. In contrast, a compressive strain increases the bandgap [16], [17].

CIGS and CdTe exhibit high efficiency, but indium and telluride scarcity is a concern for scaling up module production. Moreover, recycling of systems is complicated because of Cd toxicity. In comparison, CZTS has ideal optoelectronic properties and is made of earth-abundant, non-toxic, and low-cost constituent elements. However, to improve the device efficiency, the structural characterization, such as the identification and quantification of secondary phases, and depiction of grain structural properties, such as strain and twinning, need further work.

An additional complication arising with CZTS is that the crystallographic structures of some of the phases involved have nearly identical lattice parameters, which makes it challenging to identify and quantify the phases. For instance, ZnS with a face-centered cubic crystal structure (F-43m), and the kesterite CZTS with a tetragonal body-centered structure (I-42m), are closely related. Doubling the a, b, or c axis of the cubic ZnS structure yields a unit cell corresponding to kesterite with a small tetragonal deformation $|c/(2a) - 1| < 0.0026$ (with lattice constants from [18]).

A progress beyond the "trial-and-error" approach is vital to visualize the microstructure and local strain within the thin films in 3D and preferably also record their evolution during processing under conditions that are deemed "realistic". Such information can guide theoretical understanding and the development of models for quantitative prediction, thereby accelerating the design efforts. Moreover, physical parameters may be determined by comparing 3D models and 3D experimental data (e.g. [19]). However, the techniques currently employed to characterize the structural properties and the local stress have significant limitations:

- *Electron microscopy* (EM) can provide atomic-scale insight [13] but is confined to studies at the surface or films of a few hundred nanometer thickness. Three-dimensional resolved mapping may be accomplished by a combination of Electron Backscattered Diffraction (EBSD) and serial sectioning using either FIB [20] or laser ablation [21]. However, the destructive procedure prohibits studies of dynamics and direct coupling to functionality. Moreover, the angular resolution achieved by EM makes quantitative stress determination difficult and does not allow for a distinction between the phases mentioned above with nearly identical lattice parameters.

- *X-ray powder diffraction* (XRD) and *grazing incidence X-ray diffraction* (GIXRD) provide bulk information about phases, orientations, and strain, but only about average properties. Typically these techniques can identify secondary phases at the level of a volume percentage of 1, but the lack of well-separated peaks in the powder diffraction patterns imply that, e.g. quantification of ZnS is not possible [22].

- *X-ray nanoprobe* and forward scattering ptychography methods relying on the use of a synchrotron can reveal the local elemental composition in 3D [23] but does not provide structural information. Moreover, sample must be quite small (<10 µm), and dynamics studies representing bulk conditions are, therefore, excluded.

- *Spectroscopic methods* like X-ray Fluorescence Spectroscopy (XRF), X-ray Absorption Near-Edge Structure analysis (XANES) [24] can reveal the elemental composition but does not reveal anything about the microstructure of the film.

In this paper, we propose three-dimensional X-ray diffraction (3DXRD) as a tool for studying the microstructure and local stress in the photovoltaic polycrystalline films. This non-destructive technique combines highly penetrating hard X-rays from a synchrotron source and the application of 'tomographic' approach to the acquisition of diffraction data [25]–[32]. For grains with a known phase and sizes larger than a few micrometers, 3DXRD can generate 3D maps of several thousands of grains, revealing their shape, orientation, and type II stress (as averaged over each grain) and their variation with time [19]. For grains with a size in the 0.1-1 µm range – as is typical for photovoltaic polycrystals – shape information is not available, but one can still determine the position, size, orientation, and strains of each grain as a function of time, and thereby generate statistics on the dynamics at the grain scale [27], [32]–[35].

However, the application of standard 3DXRD software to thin-film solar cells is hampered by the complication of phase identification. In principle, a standard single crystal crystallographic analysis can be applied to each grain, a method known as multigrain crystallography [29], [36]. Here we present an approach where *a priori* information about the photovoltaic materials is used to facilitate the generation of comprehensive statistics of phase, grain size, strain, and twinning relations by standard 3DXRD software. We discuss the importance of such data for R&D in photovoltaics and outline how this work can be generalized to the generation of 3D *in situ* movies of the microstructure.

The method will be presented with reference to and demonstrated on a specific example: a CZTS (kesterite) solar cell device. We examine the crystallographic properties of this semiconductor on the grain level and the mechanical deformation in the film that the experimental data reveal. Moreover, we present approaches to get around the crystallographic challenges that this absorber layer imposes in order to identify and quantify secondary phases, stress values, and twin boundaries in the material. In our view, other chalcogenide thin-film systems such as CIGS and CdTe could also benefit from this type of 3DXRD analysis gathering statistical information about the absorber film microstructure buried in the multilayer device structure.

In the first part of this study, we present the crystallography related to the CZTS absorber layer and the challenges of identifying the secondary phases. Next, we introduce the principles of the 3DXRD technique within the context of the absorber layer microstructure and present an appropriate data analysis pipeline. Subsequently, we present an experimental 3DXRD study of CZTS including the sample details, and the results. In the final part, we discuss the connection of the results to photovoltaics properties and how recent developments of 3DXRD can advance the characterization of thin films even further.

## 3   Crystallographic aspects of kesterite

First, we must distinguish between kesterite and disordered kesterite, the latter the most frequently observed structures for CZTS. X-ray and Neutron studies have demonstrated that the quaternary compound CZTS, crystallizes in the kesterite structure (I-4) [37],[38]. The "disordered" kesterite structure associated with space group (I-42m ) was first observed by *Schorr* [39]. In this phase Cu and Zn cations intermix in the Cu-Zn layers (z=1/4, 3/4) of stoichiometric CZTS [39], see Figure 1a. The critical temperature for the phase transition from the ordered to disordered kesterite is reported to be in the range  $T_c$ = 480-560 K [40],[18],[41]. These temperatures are all below the annealing treatments under which CZTS is usually grown (720-830 K). Therefore, disordered kesterite will form initially, and ordering among Cu and Zn can only be controlled during the cooling process.

Moreover, CZTS films are grown in Cu-poor, Zn-rich conditions to obtain high-efficiency devices [42], [43]. The off-stoichiometric CZTS maintains the kesterite-type structure with variations in the lattice parameters due to the altering composition and the cation disorder [44]. The pure-phase kesterite phase only exists in a narrow area of the $ZnS-CuS_2-SnS_2$ phase diagram [45], [46]. Thus, secondary phases tend to form in the off-stoichiometric films, for instance, ZnS with a face-centered cubic crystal structure (F-43m), Figure 1 b). The two structures are closely related: doubling the a, b, or c axis of the cubic structure of the ZnS yields a unit cell corresponding to kesterite and with nearly identical lattice parameters, see Table 1.

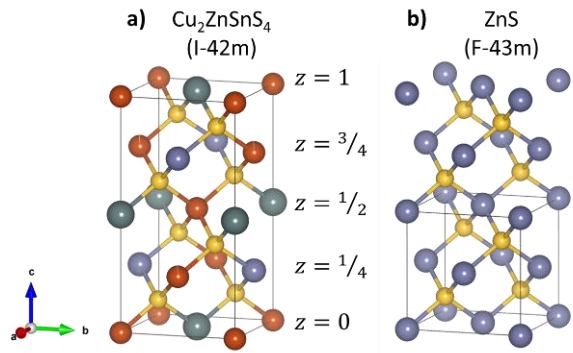

*Figure 1. a) Kesterite crystal structure. b) Sphalerite crystal structure (two unit cells are depicted, only one is marked). (orange: Cu, lilla: Zn, gray: Sn, yellow: S). The crystal structures were drawn with the VESTA computer program* [47].

*Table 1 Lattice parameters of ZnS and CZTS.*

| Phase | Space group | Unit cell parameters | ICSD No. | Reference |
|---|---|---|---|---|
| ZnS | F-43m | a, b, c = 5.4340 Å<br>α, β, γ = 90º | 77090 | [48] |
| | | a, b, c = 5.4032 Å<br>α, β, γ = 90º | 230703 | [49] |
| Cu$_2$ZnSnS$_4$ | I-4 | a, b= 5.4337 Å<br>c= 10.8392 Å<br>α, β, γ = 90º | 239674 | [18] |
| | I-42m | a, b= 5.4326 Å<br>c= 10.8445 Å<br>α, β, γ = 90º | 239684 | |

## 4 3DXRD methodology

### 4.1 3DXRD geometry and formalism

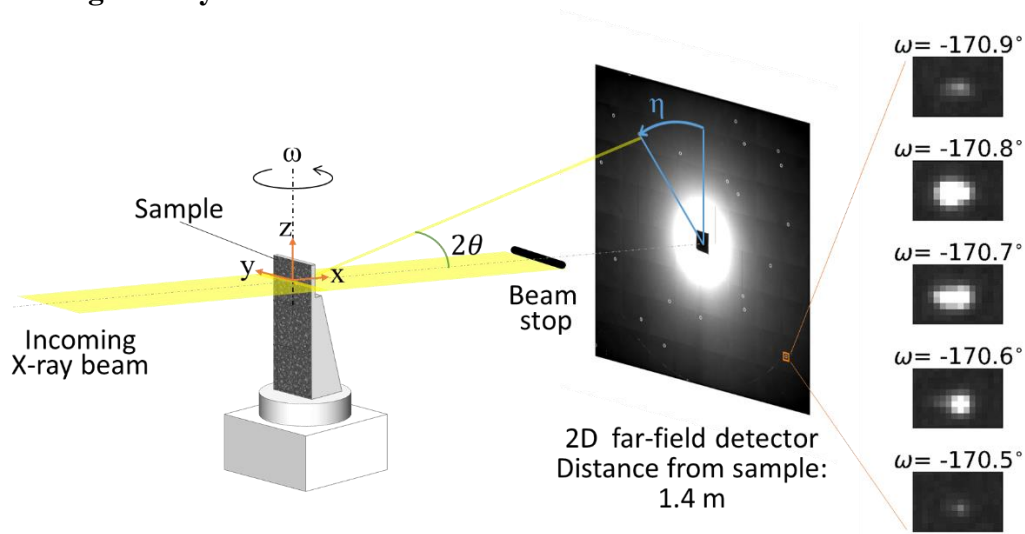

*Figure 2. Sketch of the 3DXRD experimental geometry. The laboratory coordinate system is defined. The diffracted beam for a reflection from some grain is characterized by the rotation angle ω, the Bragg diffraction angle θ, and the azimuthal angle η. The evolution of a diffraction spot associated with a given grain reflection, framed in an orange box on the detector, is shown as function of ω: it appears at -170.9° and disappears at -170.5°.*

3DXRD is a well-established tool for non-destructive characterization of grains in 3D. Based on the use of a monochromatic beam from a synchrotron source, the experimental geometry is sketched in Figure 2. Similar to the rotation method, diffraction images are acquired while rotating the sample around an axis (ω) perpendicular to the incoming beam. It images the intensity of the diffraction spots originating from the individual grains. Figure 2 displays a stack of recorded diffraction images for a small rotation range, showing the evolution of the intensity within a region of interest comprising the diffraction spot from one reflection. Typically, a focused line beam is used, thereby characterizing one slice in the sample. To provide 3D information, the sample is then and translated along z, and the data acquisition is repeated. In this way, one characterizes multiple slices that correspond to consecutive z-positions in the sample. In the far-field version of 3DXRD, which is of interest here, the sample-detector distance is relatively large (tens of centimeters to meters), and the size of the detector pixels (a few hundred µm) similar to the size of the sample. This geometry is optimized for high angular resolution.

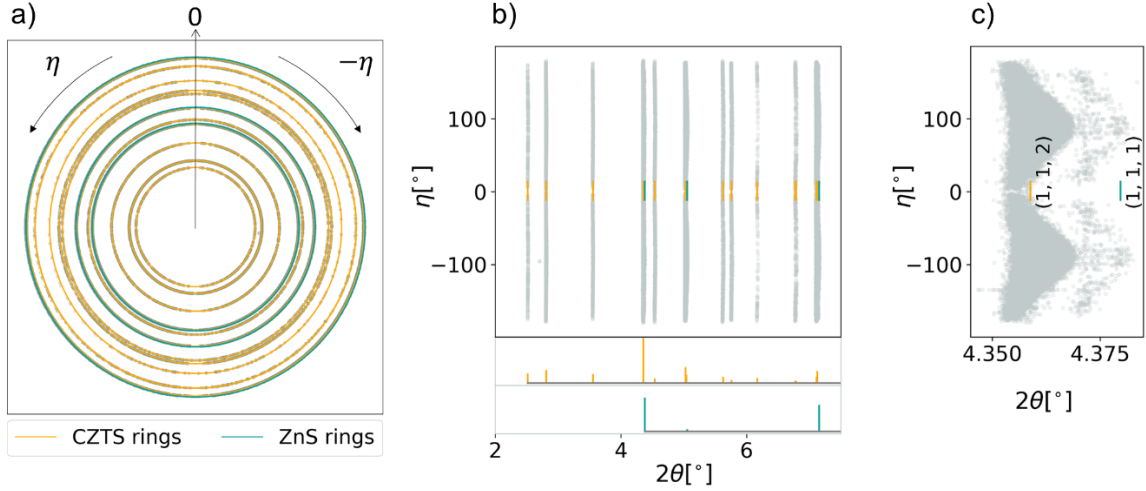

*Figure 3. a) Plot of the position of the diffraction spots on the detector (sum over all ω) and b) corresponding plot of the azimuthal angle, η, vs. the two-theta angle (2θ) of the diffraction spots. In both cases the lines related to the two phases CZTS and ZnS are identified. c) Zoom in on the overlapping CZTS (112) ring and the ZnS (111) ring.*

For the relation between experimental observables (position of diffraction peaks on the detector and corresponding rotation angle ω) and reciprocal space, we shall follow the FABLE conventions [50]. Figure 3 shows a plot of the position of all the harvested reflections from all the slices and their azimuthal angle η, according to the FABLE protocols. Let $\underline{G}_l$ be the reciprocal lattice vector corresponding to lattice planes (h,k,l) in a particular grain of interest, as defined in the laboratory system (see Figure 2). Let $\underline{G}_s$ be the same diffraction vector defined in the sample system – fixed with respect to the sample. The diffraction condition is fulfilled when

$$\vec{G}_l = \boldsymbol{\Omega}\vec{G}_s = \boldsymbol{\Omega UB}\begin{pmatrix}h\\k\\l\end{pmatrix} = |\vec{G}|\begin{pmatrix}-\sin(\theta)\\-\cos(\theta)\sin(\eta)\\\cos(\theta)\cos(\eta)\end{pmatrix}. \tag{1}$$

Here $\boldsymbol{\Omega}$ is the rotation matrix corresponding to rotation around the ω-axis, **U** is a matrix representing the orientation of a grain of interest, (θ, η) are polar coordinates characterizing the direction of the diffracted beam, see Figure 2, and $|\vec{G}| = \frac{2\sin(\theta)}{\lambda} = \frac{1}{d}$ is given from Bragg's law, with d representing the spacing between crystallographic planes. **B** is a matrix that comprises information about the unit cell as expressed by reciprocal lattice constants ($a^*, b^*, c^*, \alpha^*, \beta^*, \gamma^*$):

$$\boldsymbol{B} = \begin{pmatrix} a^* & b^*\cos(\gamma^*) & c^*\cos(\beta^*) \\ 0 & b^*\sin(\gamma^*) & -c^*\sin(\beta^*)\cos(\alpha) \\ 0 & 0 & c^*\sin(\beta^*)\sin(\alpha) \end{pmatrix} \text{ with } \cos(\alpha) = \frac{\cos(\beta^*)\cos(\gamma^*)-\cos(\alpha^*)}{\sin(\beta^*)\sin(\gamma^*)}. \quad (2)$$

As usual in crystallography, the matrix $\mathbf{A} = (\mathbf{B}^{-1})^T$, where $(\ldots)^T$ symbolizes transposing, comprises the corresponding information about the direct space unit cell of the grain of interest, expressed by the direct space lattice constants (a, b, c, α, β, γ). Notably, the matrix inverse of $(\mathbf{UB})^{-1}$ gives the real space unit cell vectors ($\vec{a}, \vec{b}, \vec{c}$) of the grain in the sample frame.

The grain elastic strain can be expressed in terms of the unit cell of a reference (unstrained) crystal $\mathbf{A}_0$ and a strained crystal **A**. We determine the deformation gradient tensor of the grain as:

$$\boldsymbol{F}^g = \boldsymbol{A}\boldsymbol{A_0}^{-1} = (\boldsymbol{B_0}\boldsymbol{B}^{-1})^T. \quad (3)$$

For the small strain levels of relevance to this study and in the absence of rotation, the infinitesimal strain tensor **ε** is applicable and is, by definition, given by the symmetric tensor

$$\varepsilon_{ij} = \frac{1}{2}\left(F^g{}_{ij} + F^g{}_{ji}\right) - I_{ij} \quad (4)$$

where **I** is the identity matrix. Ultimately the strain is transformed in the sample coordinate system by applying the orientation of the grain:

$$\boldsymbol{\varepsilon}_{sample} = \boldsymbol{U} \cdot \boldsymbol{\varepsilon}_{grain} \cdot \boldsymbol{U}^T \quad (5)$$

### 4.2 Conventional 3DXRD data analysis and its relation to polycrystalline photovoltaic materials

3DXRD methods are usually applied to studies of polycrystalline materials. It requires knowing the space group and unit cell lattice parameters for the unstrained material (that is, with a known matrix $\mathbf{B}_0$). The position of diffraction spots on the detector are given by the grain orientation with only small perturbations due to strain. In this case, one may initially assume that all grains are associated with the undistorted matrix $\mathbf{B}_0$. The process of identifying grains, multigrain indexing, then becomes a question of identifying orientations, **U**, that complies with Eq.(1), for a set of known (h,k,l) indices. As a result, the reflections determined are sorted into groups, where each group represents one grain. The main limitation is the overlap of diffraction spots. For inorganic materials exhibiting weak textures, up to around 3000 grains can be indexed [28] from a single rotation scan. Our approach is to utilize a line beam that limits the number of simultaneously illuminated grains to avoid spot overlaps.

Following this indexing step, all the tools of single-crystal crystallography can be applied to each grain. The relative grain volume can be estimated from the integrated intensities of the assigned reflections. A least-square fit can be performed to determine all nine **U** and **B** components, by minimizing the angular distance between the predicted reflections, cf. Eq. (1), and the experimentally determined ones. Next the strain tensor can be determined from **B** by Eqs. (3)-(4), where the grain unit cell, determined during indexing, is compared to an unstrained reference **B$_0$** [26]. When required, crystal structure refinement can also be used to optimize the position of the atoms within the unit cell – with a quality in the results that can match that of refinement based on single-crystal diffraction [51], [28].

In principle, the 3DXRD formalism, as expressed by Eq. (1), allows for indexing without any *prior* information by operating in the 9D space, spanned by **U** and **B**. In this way, 3DXRD could handle any number of arbitrary unknown phases, strained or unstrained. However, brute force procedures are too slow to be operational. A general-purpose method involving searching only in 3D has been suggested [36], but this algorithm still lacks a sufficiently robust software implementation. In this work, phase identification from a database search could provide sufficiently accurate for the unit cell parameters of the phases in the sample.

The polycrystalline photovoltaic materials, and particularly the kesterite solar cell, pose a special challenge as several complications are present simultaneously:

- Several phases are present, and some might not be known *a priori*.
- Some phases may exhibit a doubling of the unit cell in one direction, and their lattice parameters give rise to 2θ-angles that are nearly indistinguishable.
- Twins may appear, leading to a large fraction of reflections being shared by more than one grain.
- The specimens are subject to mechanical stress, originating in the thin film preparation.
- The grains are sub-micron in size leading to signal to noise ratios (S/N) in the diffraction data.

These CZTS data comprise additional information, as the doubling of a unit cell leads to superstructure peaks. These may, however, be weak, and spurious peaks from other phases can cause unexpected overlaps.

In the following, we present an approach that overcomes these challenges and generates a list of grains. Each grain is associated with an orientation, a size, and a phase related to a unit cell. The unit cell parameters represent a strained state, caused by stoichiometry changes and an externally imposed mechanical strain. We demonstrate how to calculate an overall strain for the film and subtract it from the oriented unit cells of the grains.

In section 4.3, we present our approach to indexing the grains and identifying their unstrained unit cells. Section 4.4 describes how to exploit the results for the statistical description of phases, grains size distributions, stresses, and texture with a special focus on potential twin relations.

In an initial exploratory phase, we discovered that a data analysis based on the existence of two phases, a cubic and a tetragonal, was consistent with the data. Figure 3b displays the lines associated with the cubic and tetragonal phases. Hence, we shall assume two phases in the following.

## 4.3 Identifying grains and their crystal structure.

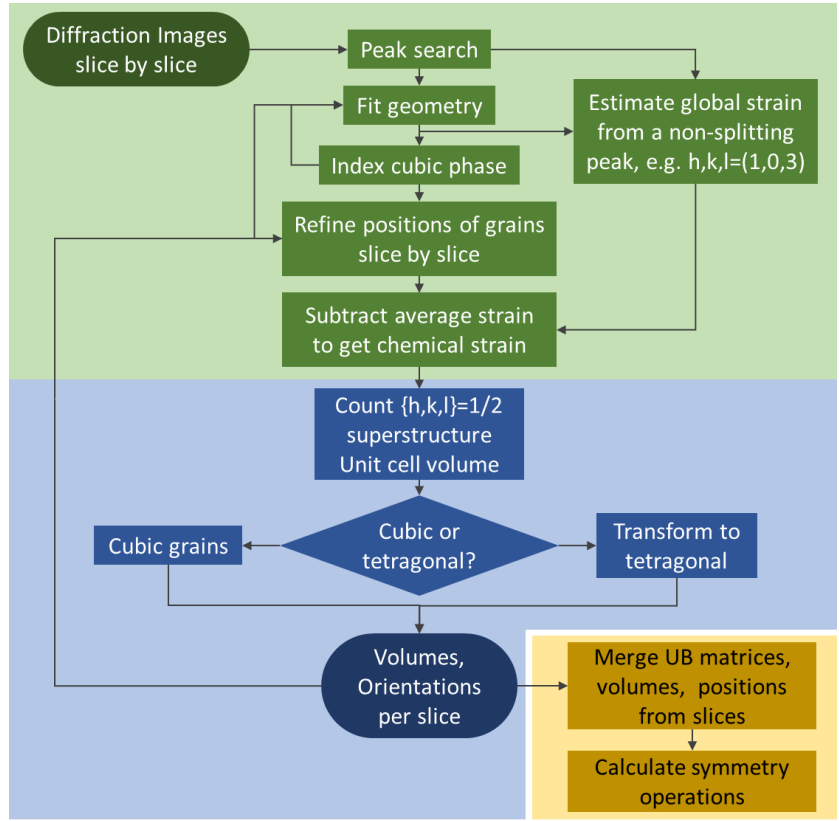

*Figure 4. Flow diagram of the data analysis pipeline for indexing grains and identifying their unstrained unit cells.*

The data analysis pipeline used is sketched in Figure 4. It is structured in three parts. Its implementation is based on existing 3DXRD software, throughout, primarily the ImageD11 software [52].

*In the first part,* the experimental data from different slices of the sample (different z-positions) are analyzed independently. For each slice, initially, a background is subtracted from the raw images, and the diffraction spots are identified ("peak search"). Based on the statistics of these reflections, several global parameters related to the experimental setup are refined, including wavelength, sample-to-detector distance, and tilts of the detector. We assume all grains belong to the same phase with a cubic symmetry and an "average" lattice constant, $a_0^*$, corresponding to a common $\mathbf{B}_0 = a_0^* \mathbf{I}$. Excluding superstructure peaks and using only diffraction spots positioned at $2\theta$ angles, corresponding to the cubic phase (the three green lines marked in Fig 3b), grains are found by the classical monophase 3DXRD indexing program ImageD11. The result is a list of grains, each with an associated $(\mathbf{UB})^{-1}$ matrix and a list of reflections.

Next, we assume that the mechanical stress gives rise to comparable strains in the grains if asserted in the sample coordinate system. The grain strain can then be expressed in terms of an "average" contribution and a residual that is specific to the grain. The average strain tensor of the film may be determined in several ways. The approach used here is to focus on the diffraction spots belonging to a specific (hkl) family. For each diffraction spot, $i$, we can determine the shift in 2θ position, $\Delta 2\theta^i_{hkl}$. The corresponding normalized scattering vector in the sample system is $\vec{n}_i = \vec{G_s^i}/|G_s|$. From the differentiation of Bragg's law, the 2θ shift corresponds to an axial strain (a strain in the direction $\vec{n}_i$) of

$$\varepsilon^i{}_{hkl} = \frac{\Delta d_{hkl}}{d^0{}_{hkl}} = -\Delta\theta^i_{hkl} \cot\theta^0{}_{hkl} \tag{6}$$

Here $2\theta^0{}_{hkl}$ is the average two-theta angle of the (hkl) Debye-Scherrer ring. By definition,

$$\varepsilon^i{}_{hkl} = \begin{pmatrix} n^i_x & n^i_y & n^i_z \end{pmatrix} \begin{pmatrix} \varepsilon_{11} & \varepsilon_{12} & \varepsilon_{13} \\ \varepsilon_{12} & \varepsilon_{22} & \varepsilon_{23} \\ \varepsilon_{13} & \varepsilon_{23} & \varepsilon_{33} \end{pmatrix} \begin{pmatrix} n^i_x \\ n^i_y \\ n^i_z \end{pmatrix}. \tag{7}$$

From this follows that the average strain tensor elements for the hkl ring, $\varepsilon_{ij}$, can be determined by a linear least-squares fit of experimental data to Eqs. (6) and (7). Let the resulting matrix be $\boldsymbol{\varepsilon}^{mat}$. Subsequently, for each grain, $\boldsymbol{\varepsilon}^{mat}$ is subtracted to correct the unit cell parameters and obtain the "strain-free" lattice parameters.

$$(\boldsymbol{UB})_0 = (\boldsymbol{I} + \boldsymbol{\varepsilon}^{mat}) \cdot \boldsymbol{UB} \tag{8}$$

Next, the grain strain tensor is calculated applying Eq. (4) and the obtained "strain-free" unit cell as the unstrained reference. Then, we calculate the stress tensor using Hook's law

$$\sigma_{ij} = C_{ijkl}\varepsilon_{hkl}. \tag{9}$$

Subsequently, the stress in the film is obtained by transforming the grain into the sample reference system.

$$\boldsymbol{\sigma}_{sample} = \boldsymbol{U} \cdot \boldsymbol{\sigma}_{grain} \cdot \boldsymbol{U}^T \tag{10}$$

From here on, we will adopt Voigt's notation to simplify the index of the tensor components, where index ij=[11 22 33 23 13 12] becomes i=[1 2 3 4 5 6]. We have chosen the elastic constants given in [17] because the converged lattice parameters agree with our experimental data. Other calculations report similar numbers [16], [53]–[56].

*In the second* part, the superstructure peaks are taken into consideration. These appear at positions in the reciprocal space that corresponds to a doubling of the direct space unit cell. To study this systematically, for each grain, we form the supercell $(2\vec{a}, 2\vec{b}, 2\vec{c})$ in direct space, a doubling in all directions of the unit cell $(\vec{a}, \vec{b}, \vec{c})$. The reciprocal space unit cell is correspondingly halved in all directions. A search is now performed within the full set of reflections from the original peak search of reflections positioned at the nodal points of the supercell. The reflections appearing with an odd number in any of the three directions are "superstructure peaks"; they do not belong to the original cell. Searching for grains with a double unit cell in the $\vec{a}$ direction, the number of reflections appearing at (h, 2k, 2l) with h odd is compared to the number of reflections assigned to the supercell. If the ratio is above a certain threshold, defined by S/N and spurious background, the grain is defined to have a double cell with the preferred axis along $\vec{a}$. Likewise, searches are performed on unit cells that are doubled along $\vec{b}$ or $\vec{c}$. It is generally observed that the shortest unit cell lattice and the axis with odd reflections coincide. No occurrence is found of cells being doubled in more than one direction.

Figure 5 shows a plot of the ratio of the number of superstructure diffraction spots to the total number of diffraction spots for a given grain as a function of the derived grain volume. For large grains, most the weak diffraction spots at low scattering angles from the superstructure are recorded. Hence, it is straightforward to classify a grain as tetragonal or cubic based on the existence of superstructure peaks. As expected, with a decreasing grain volume, more of the superstructure peaks become too faint to be recorded. For relatively small grains with few or no superstructure reflections, the distinction between tetragonal or cubic was therefore based on the unit cell geometry. This classification scheme turns out to be very robust.

For grains that are classified as tetragonal, the shortest axis is doubled. Moreover, the unit cell axes were permuted, such that this short axis becomes the $\vec{c}$-axis.

The (illuminated) volume of the grain, assuming proportionality of the volume with the reflection intensities, is given by Eq. (11), where $\sum \bar{I}_g$ is the sum of the average intensities of all indexed grains, and $V_{sample}$ is the illuminated sample volume.

$$V_{grain} = \frac{\bar{I}_{grain}}{\sum \bar{I}_g} \times V_{sample} \tag{11}$$

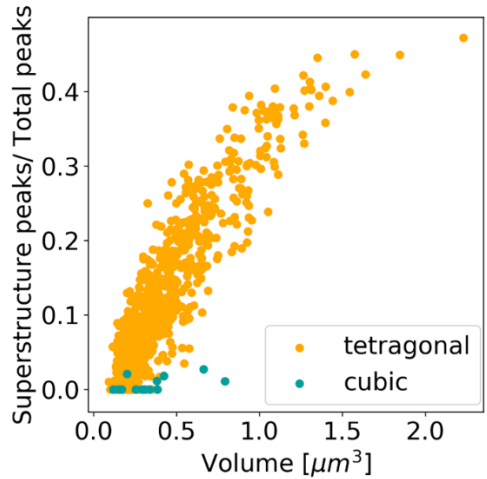

*Figure 5. The ratio of the superstructure peaks to the total number of peaks for a given grain is compared to the grain volume.*

*In the third part* of the flow diagram, the slices are combined. Grains in neighboring slices with a similar $(\mathbf{UB})^{-1}$ within given tolerances are considered identical. For these, the weighted average of $(\mathbf{UB})^{-1}$ by the integrated intensities is determined. Moreover, the (sub) volumes determined in the different slices are added. Following this, for each grain, a final optimization step is performed. Here all parameters in $(\mathbf{UB})^{-1}$ and the 3D center-of-mass position of the grain (which will cause minor translations of diffraction peaks not described by Eq. 1) are refined.

The result of the data analysis is, therefore, a division of the grains into two "phases": one with no cell doubling and with a unit cell that is close to cubic, and another with a unit cell that is close to a tetragonal unit cell with unit cell parameters (a, a, 2a, 90º, 90º, 90º). We shall term these the "cubic" phase associated with ZnS and the "tetragonal phase" associated with CZTS. Each grain is associated with an orientation, **U**, a **B** matrix, a center-of-mass position, and a list of reflections. Moreover, we have determined the average strain tensor within the (illuminated part of the) sample.

### 4.4 Statistics on the structural and mechanical properties of the grain ensemble

As already stated, once the grains have been identified, their properties can be subject to statistical analysis for understanding both average properties and the heterogeneity within the sample. The local texture (within the volume studied) can be derived from the grain orientations. The grain sizes and their distribution can be determined from the integrated intensities. Strain components and their distributions can be derived via Eqs. (3)-(4), changes in stoichiometry may be inferred directly from changes in unit cell parameters.

The spatial resolution of far-field 3DXRD did not allow us to identify neighboring grains, and therefore their misorientation angle is not accessible. But *twins* can be identified from angular relationships. In practice, reflections can be shared among grains because of coincidental overlap or due to the presence of twin relationships between grains, so thresholds must be introduced in the data analysis. Notably, twin relationships may also exist between the tetragonal and the cubic phase.

We calculate the number of shared reflections between grains by computing the scattering vectors of each grain to look for overlapping of Bragg peaks. Using equation (1), the error of the computed hkl should be below ~0.02 to be considered as part of the grain. We confirm the twin relation among pairs with 30% of reflections overlapping, if the pair has a certain misorientation angle associated to a symmetry operation.

When comparing a pair of grains, we compute a natural lattice mismatch via the deformation gradient tensor $\boldsymbol{F}$. Based on Eq. (3), here we utilize the reciprocal lattice of a grain $(\boldsymbol{UB})_n$, considered as the reference lattice that has been deformed by the inverse transpose of $\boldsymbol{F}$ when compared to grain m, whose reciprocal lattice is $(\boldsymbol{UB})_m$.

$$\boldsymbol{F} = (\boldsymbol{UB})_m^{-T} \cdot (\boldsymbol{UB})_n^{T} \qquad (12)$$

Following conventions in the field of continuum mechanics, we perform a polar decomposition of **F** to produce a pure rotation **R** and a pure stretch tensor **U**$_s$, the right stretch tensor (not to be confused with **U**-rotation matrix in 3DXRD)

$$\boldsymbol{F} = \boldsymbol{R} \cdot \boldsymbol{U}_s \tag{13}$$

Next we calculate the Biot strain tensor Eq. (14), which is equivalent to the infinitesimal strain tensor in the absence of rotations. The overall strain magnitude is given by the Euclidean norm of the Biot strain tensor $\|E\|$.

$$\boldsymbol{E} = \boldsymbol{U}_s - \boldsymbol{I} \tag{14}$$

This rotation matrix **R**, is used to determine the angle and axis of rotation between the two grains being compared. The equivalent symmetry operations are applied to the reference grain, calculating the strain magnitude. The symmetry transformation with the lowest strain value describes the misorientation angle between two grains, where the lattices are also well matched. For the tetragonal structure, eight transformations are possible, whereas the cubic structure allows 24.

## 5 Experimental

### 5.1 Sample description and preparation

The investigated kesterite solar cell architecture consists of a stack of layers deposited on a molybdenum coated soda-lime glass substrate (Mo-SLG). The CZTS absorber layer is fabricated by pulsed laser deposition (PLD) and annealed in a high-temperature sulfurized atmosphere to form the polycrystalline kesterite film (~400 nm thick). The subsequent coatings are a CdS buffer layer (60 nm), an intrinsic ZnO window layer (50 nm), an Indium Tin Oxide (ITO) contact layer (200 nm), and an $MgF_2$ anti-reflection coating (100 nm). The fabrication details can be found in [57].

We cut a 40 (W) x 300 (H) µm² piece of the solar cell, as shown schematically in Figure 6a. To maximize the signal to noise ratio of the diffracted intensity originating from the 1 µm sized grains, we reduced the 1 mm thickness of the Mo-SLG by mechanical polishing and milling by a focused ion beam (FIB) down to 4 µm thickness (see Figure 6b).

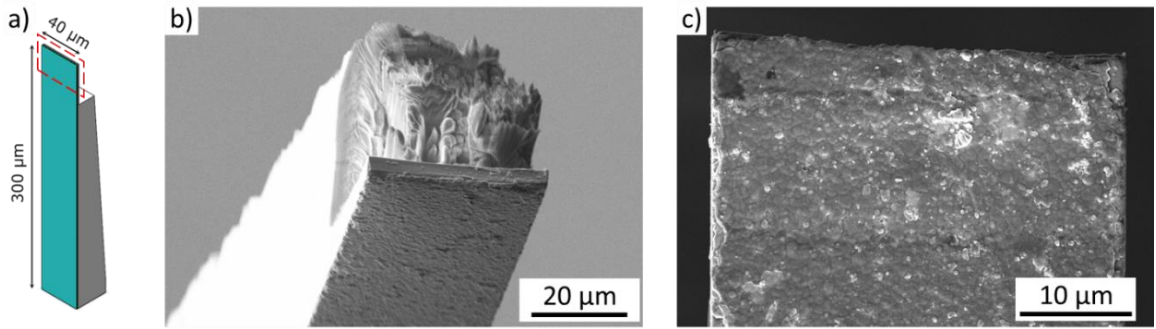

*Figure 6 a) Dimensions of the cut piece of the kesterite CZTS solar cell. The red box is where the scans of the slices are measured. b) SEM image of the tip of the sample. c) Front view of the solar cell.*

## 5.2 3DXRD experiment

The experiment was carried out at the Advanced Photon Source synchrotron at the 1-ID beamline. A monochromatic X-ray beam (52 keV) focused to a size of 1.5 µm (FWHM) (V) x 200 µm (H) illuminated the solar cell parallel to its normal plane. The range of oscillation was $\omega \in [-180°, 180°]$ with a step size of $\Delta\omega=0.1°$. A GE Revolution 41RT flat-panel detector (2048 x 2048 pixels, 200 µm pixel pitch) recorded the diffraction images. The acquisition time per slice was about 1.4 hours with 1.2 s per frame. The translation step along the z-axis was 1 µm, capturing an overlap of 0.5 µm between slices. A standard $LaB_6$ powder (NIST SRM 660c) was used for the initial calibration of the geometry of the set-up.

## 6  Results

Following the procedure outlined above, a total of 597 grains were identified, 582 tetragonal and 15 cubic. As a figure-of-merit we note that 33% of the diffraction spots identified by the peak search were assigned to these grains. It is possible to identify more grains, but they will be associated with larger errors.

### 6.1  Averaged strain and stress in the sample and the grains.

Shown in Fig 7a is the variation of the 2θ angle with the azimuthal angle, η, and with the rotation angle, ω, for all reflections in the (103) lattice plane. We define the reference angle $2\theta^0{}_{103}$ as the average angle. There is a systematic displacement of the 2θ angle with both η and ω, which we attribute to an external stress field. Hence, we can determine a strain tensor, corresponding to this external stress, using the fitting procedure formulated in Eqs. (6)-(7).

The average strain tensor elements for each of the 7 slices obtained by this least-square approximation are listed in Table 2. By subtracting the strain in the tetragonal and cubic grains, we can observe the improvement of the lattice parameters in both phases, as the distributions of the corrected lattice parameters become narrower (see Figure 7b, c).

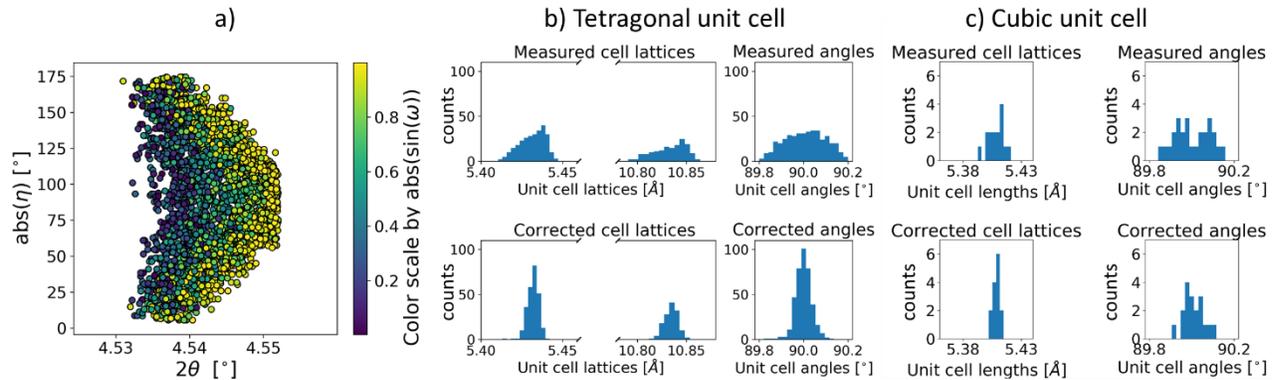

*Figure 7. a) Absolute azimuthal angle vs. 2θ angle of the diffraction spots in the (103) plane; the colors symbolize the corresponding ω-position. b) Distribution of lattice parameters for the tetragonal grains; c) Distribution of lattice parameters for the cubic grains. In both b) and c) the term the "measured" denotes the original data while the term "corrected" are the results after subtracting the effect of an external stress acting on the film.*

Subsequently, for each grain we can derive its strain tensor elements using the corrected unit cells as the reference unstrained state and Eqs. (3)-(4). Figure 8a defines the elements of the strain tensor in the tetragonal crystal structure. The normal strains directions of $\varepsilon_1$, $\varepsilon_2$, $\varepsilon_3$ are perpendicular to the (100), (010), (001) planes, whereas the shear strains $\varepsilon_4$, $\varepsilon_5$, and $\varepsilon_6$ are coplanar to the planes where the normal strain is applied. The shear strains are in equilibrium, implying that $\varepsilon_4$: $\varepsilon_{23} = \varepsilon_{32}$, $\varepsilon_5$: $\varepsilon_{13} = \varepsilon_{31}$, $\varepsilon_6$: $\varepsilon_{12} = \varepsilon_{21}$, and therefore they appear in two planes. Figure 8b shows the resulting correlation between the normal strain components $\varepsilon_1$ and $\varepsilon_3$. We observe the typical behavior of a deformed object where an increasing strain along the a-lattice will result in a decreasing strain along the c-lattice. The slope is determined to be -0.83 by a linear fit to the data. However, we also note a substantial scatter in these data, caused by grain-grain interactions.

Next, we use Eq. (9) and the corresponding elastic constants to calculate the stress components for each grain. Histograms of these components are shown in Fig 8c for the tetragonal phase. The normal stress along the a-direction $\sigma_1$ has a slightly right-skewed distribution suggesting predominant compressive stresses, whereas the normal stresses along b ($\sigma_2$) and c ($\sigma_3$) directions have left-skewed distributions that correspond to tensile stresses. The shear stress $\sigma_4$ has an almost normal distribution, whereas $\sigma_5$ has a bimodal distribution, and $\sigma_6$ a left-skewed one.

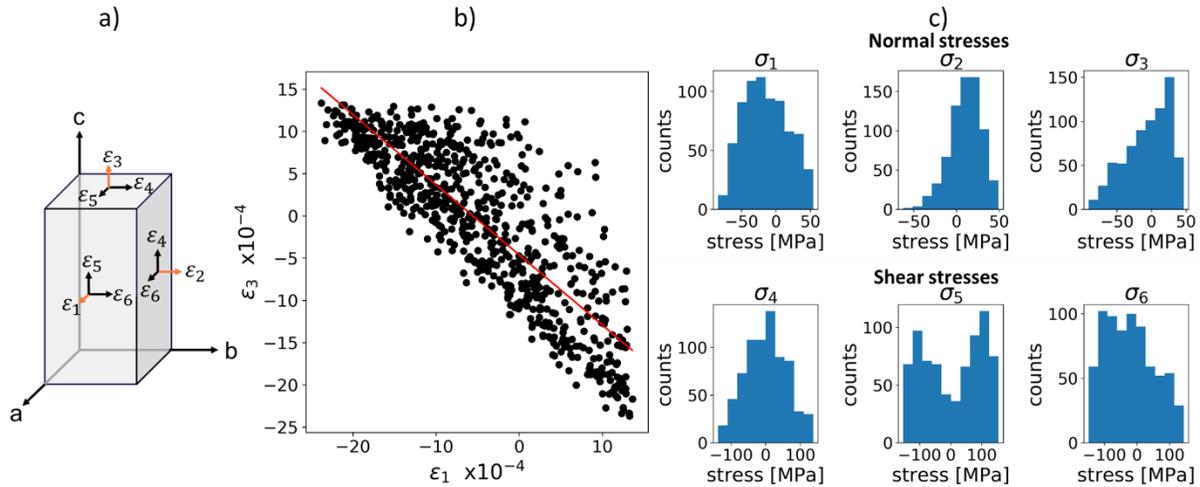

*Figure 8 a) Definition of strain components within the tetragonal grain unit cell. b) Plot of the normal strain $\varepsilon_3$ vs. $\varepsilon_1$ with a best fit to a linear regression. c) Histograms of the corresponding stress components in the grain unit cell. All the retrieved tetragonal grains are represented in b and c.*

We can now calculate the macroscopic stresses in the sample by averaging over the grains while taking into consideration their orientation. The results are listed for each slice in Table 3 and displayed in Figure 9a. We visualize the normal stresses $\sigma_1$, $\sigma_2$, and $\sigma_3$ along the x-, y-, and z-axes correspondingly. The shear stresses are coplanar to the planes where the normal stresses are applied. Similar to the strain depiction, the shear stresses appear in two planes as they are in equilibrium, meaning that $\sigma_4$: $\sigma_{23} = \sigma_{32}$, $\sigma_5$: $\sigma_{13} = \sigma_{31}$, and $\sigma_6$: $\sigma_{12} = \sigma_{21}$.

The film shows a compressive strain in the normal direction of the film plane that corresponds to an averaged compressive stress of -144.7 MPa. The corresponding tensile strain within the film plane results in an average stress of 53.6 MPa along the y-axis and 81.9 MPa along the z-axis (Figure 9). The shear strains $\varepsilon_4$ and $\varepsilon_5$ are zero within experimental error. On the other hand, the non-zero component $\varepsilon_6$, coplanar to the (xz)- and the (yz)-planes, is associated with a compressive shear stress $\sigma_6$ = -38.1 MPa. This can be explained by a misalignment of the sample in the diffractometer.

*Table 2. The $\varepsilon^{mat}$ strain tensor components per slice for the tetragonal phase obtained from the 2θ angle displacements (Eq. (7)).*

| Strain component | Slice 1 Value [× $10^{-4}$] | Slice 2 Value [× $10^{-4}$] | Slice 3 Value [× $10^{-4}$] | Slice 4 Value [× $10^{-4}$] | Slice 5 Value [× $10^{-4}$] | Slice 6 Value [× $10^{-4}$] | Slice 7 Value [× $10^{-4}$] | Average Strain [× $10^{-4}$] |
|---|---|---|---|---|---|---|---|---|
| $\varepsilon_1$ | -19.1 | -20.5 | -20.8 | -22.1 | -22.2 | -22.5 | -21.0 | -21.2 |
| $\varepsilon_2$ | 8.4 | 9.6 | 8.0 | 8.6 | 8.5 | 7.3 | 7.7 | 8.3 |
| $\varepsilon_3$ | 10.5 | 11.7 | 13.1 | 11.8 | 12.8 | 13.3 | 13.7 | 12.4 |
| $\varepsilon_4$ | -0.3 | -0.3 | -0.3 | -0.4 | -0.2 | 0.4 | 0.4 | -0.1 |
| $\varepsilon_5$ | -0.8 | 0.8 | 1.3 | 1.1 | 1.4 | -0.0 | 0.4 | 0.6 |
| $\varepsilon_6$ | -5.6 | -5.5 | -4.9 | -5.2 | -5.0 | -6.3 | -5.4 | -5.4 |

*Table 3. The average stress tensor components per slice obtained after calculating grain stresses (Eq.(10) ).*

| Stress component | Slice 1 Value [MPa] | Slice 2 Value [MPa] | Slice 3 Value [MPa] | Slice 4 Value [MPa] | Slice 5 Value [MPa] | Slice 6 Value [MPa] | Slice 7 Value [MPa] | Average Stress [MPa] |
|---|---|---|---|---|---|---|---|---|
| $\sigma_1$ | -132.5 | -134.8 | -138.0 | -156.2 | -151.6 | -156.1 | -144.0 | -144.7 |
| $\sigma_2$ | 54.1 | 68.2 | 54.8 | 50.2 | 52.2 | 38.9 | 57.0 | 53.6 |
| $\sigma_3$ | 73.4 | 82.1 | 92.0 | 72.7 | 80.4 | 79.5 | 93.3 | 81.9 |
| $\sigma_4$ | -5.7 | -4.4 | -2.6 | -3.2 | -2.9 | 1.5 | 5.0 | -1.8 |
| $\sigma_5$ | -4.5 | 7.4 | 7.6 | 6.0 | 6.7 | 1.9 | 4.1 | 4.2 |
| $\sigma_6$ | -39.1 | -38.4 | -37.4 | -35.3 | -34.0 | -44.4 | -38.4 | -38.1 |

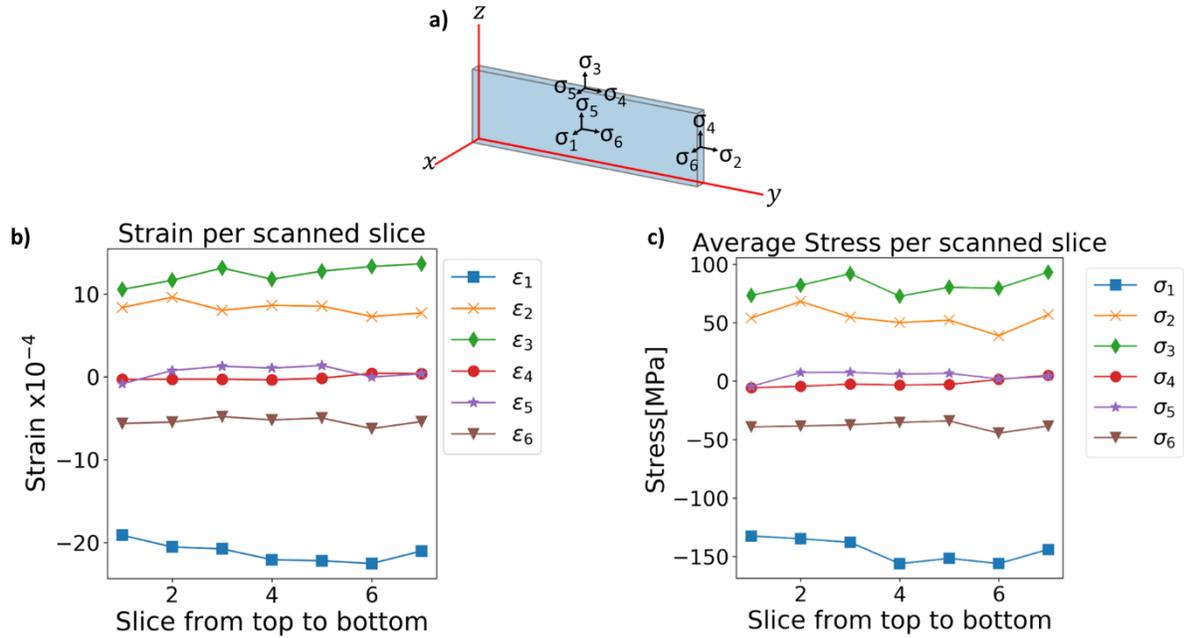

*Figure 9 a) Definition of the stress components with reference to one of the measured slices. b) strain and c) stress components, per scanned slice from the tip of the CZTS solar cell (slice 1) downwards.*

### 6.2 Grain size distribution and orientations

The grain volumes were derived using Eq. (11). The histogram for the tetragonal phase shown in Fig 10 a) and b) are consistent with a log-normal size distribution with a cut-off at lower radii due to thresholding of the intensities. The grain size is 0.32 ±0.26 µm$^3$, and the corresponding radius 0.47 ±0.18 µm (see Figure 10a, b). The radius is obtained assuming a cylindrical volume for the CZTS grain with a height of 0.45 µm, the film thickness. For the cubic grains, the grain statistics is scarce. The volume is 0.25 ±0.16 µm$^3$, Figure 10c), and the corresponding equivalent-sphere radius is 0.43 ±0.13 µm, (see Figure 10d).

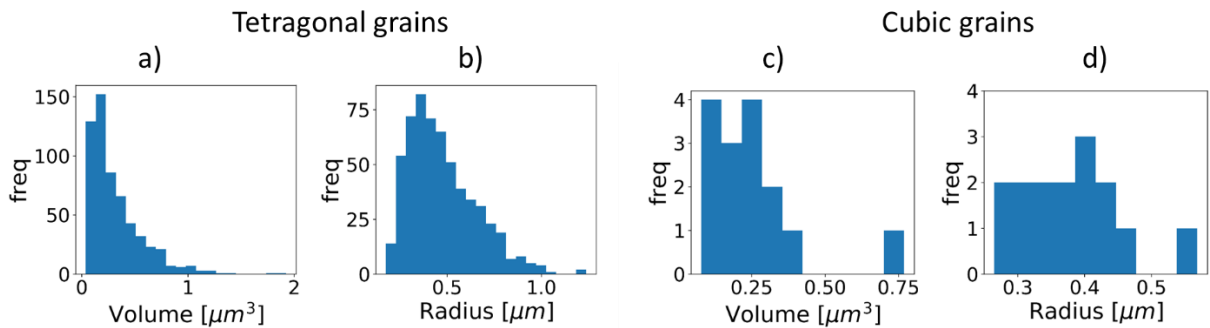

*Figure 10 Histograms for the tetragonal grains: a) volume, b) radius. Histograms for the cubic grains: c) volume, d) radius.*

The orientation distribution functions were computed using the MTEX MATLAB toolbox and a 5° resolution [58]. The pole figures of the tetragonal grains for the planes {100}, {110}, {001} and {112} are shown in Figure 11, with the yz-plane being the film plane on the diffractometer. The pole figures for the cubic grains are not shown because of poor grain statistics (15 grains).

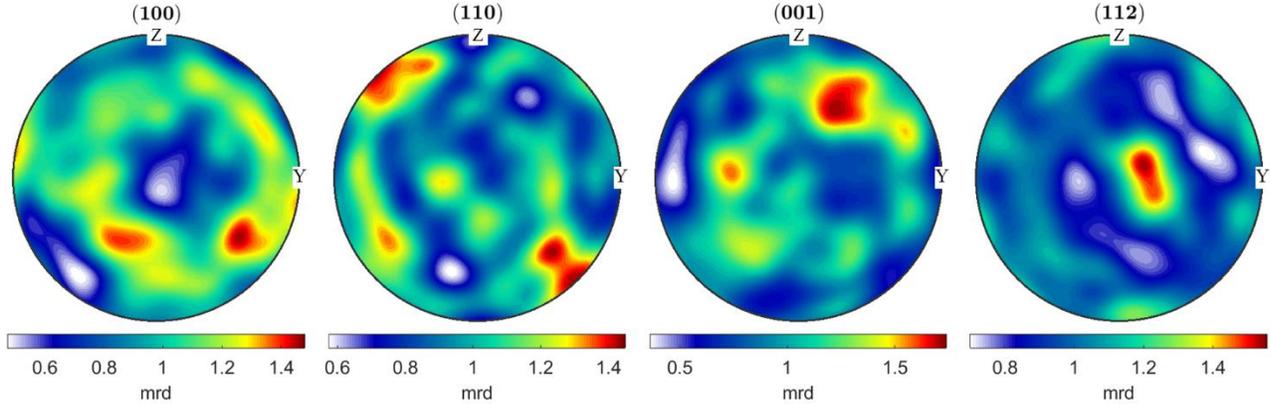

*Figure 11 Pole figures of the tetragonal grains in multiples of random distribution (mrd).*

### 6.3 Twin boundaries

Twin boundaries are often described with the quantity Σ, which is defined by the ratio between the area enclosed by a unit cell of the coincident lattice sites and the standard unit cell.

To identify twins among the tetragonal grains a search was performed for pairs of grains sharing 30% of the reflections or more. We find that the misorientation angles of these in all cases correspond to one of four values, corresponding to the symmetry operations with the minimum strain between the compared grains.

120 twin pairs were identified. Their resulting misorientation angles are shown in Figure 12a-d. The most frequent type of grain boundary is identified as Σ3, characterized by a rotation of 70.53° and the corresponding symmetrically equivalent misorientation angles 109.47°, 131.81°, and 180°. Σ3 boundaries are also detected by EBSD in chalcopyrite thin films [59]. Our measurements deviate from the mentioned angles due to the tetragonal distortion $c/2a < 1$ in the kesterite structure. We report the average rotation angles, the corresponding rotation axis and the transformed lattice plane in Table 4.

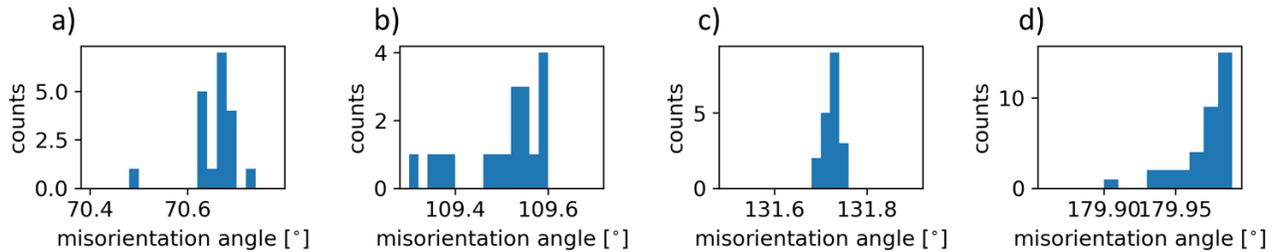

*Figure 12 Distribution of the misorientation angles for pairs of CZTS grains sharing 30% of their reflections. These fall in four groups around a) 70.66°, b) 109.50°, c) 131.73°, and d) 179.98°.*

*Table 4 Total counts of Σ3 boundaries according to the rotation angle*

| Rotation angle | Rotation axis | Transformed plane | No. twinned pairs | twinned grains/ total tetragonal grains × 100 [%] |
|---|---|---|---|---|
| 70.65° | $<110>_{tetra}$ | $\{110\}_{tetra}$ | 19 | 6.5 |
| 109.50° | $<110>_{tetra}$ | $\{110\}_{tetra}$ | 4 | 1.4 |
|  | $<201>_{tetra}$ | $\{102\}_{tetra}$ | 14 | 4.8 |
| 131.73° | $<401>_{tetra}$ | $\{101\}_{tetra}$ | 19 | 6.5 |
| 179.98° | $<201>_{tetra}$ | $\{112\}_{tetra}$ | 39 | 13.4 |
|  | $<111>_{tetra}$ | $\{114\}_{tetra}$ | 25 | 8.6 |
|  |  | **Total** | 120 | 41.2 |

Among the cubic grains, only one pair of twinned grains was found with a rotation of 180°-$<211>_{cubic}$ transforming the plane $(211)_{cubic}$. This transformation corresponds to a Σ3 twin boundary. The equivalent symmetric transformation is the Σ3 60°-$<111>_{cubic}$, typically found in the diamond-type structure[59].

Special orientation relationships between grains of the different phases were not considered.

## 7 Discussion

### 7.1 The kesterite solar cell

**Identification of secondary phases.** The 3DXRD analysis of the CZTS absorber layer revealed the presence of the ZnS phase representing 2.5% of the total number of grains in the film. This small amount of ZnS is not detrimental to CZTS devices. However, when present in large amounts, it can block the charge transport or increase the series resistance in the solar cell [60].

A similar value has been measured in a sputtered CZTS film with 3.1% of ZnS by X-ray absorption near-edge spectroscopy (XANES) at the sulfur K-edge [61]. A co-sputtered CZTS film, also measured by XANES, yielded 10.5% of ZnS [61]. These measurements are within the ZnS limit detection in XANES of 3% [24].

**Film averaged stress.** The stress and strain components vary slightly between slices with an average standard error of 2.23 MPa. These almost constant values are a testament to the robustness of the 3DXRD method.

Comparing to literature, Johnson *et al.* consider the formation of a biaxial tensile thermal stress at the Mo/CZTS interface during annealing as a result of the Mo deposition stress and the thermal expansion mismatch stresses [62]. In their study, wafer curvature measurements show compressive deposition stresses of about -400 MPa to -38MPa for optimized Mo sputtering deposition, and a compressive deposition stress of about -100 MPa for CZTS by co-sputtering. The normal tensile stresses over the PLD-CZTS film plane agree with the biaxial tensile stresses of the co-sputtered CZTS, although the compressive normal element and the shear components are missing in their model. Moreover, our calculated stresses are in the same order of magnitude as their deposition stresses.

In addition to the thermal stresses inflicted during annealing, we can also consider the entangled combination of the chemical distortion caused by the off-stoichiometric composition of the film and the inflicted mechanical deformation through the cutting of the solar cell and the removal of the glass through mechanical polishing. Likewise, FIB etching could have introduced artifacts such as ion implantation and structural damage [63], [64]. Unfortunately, sample preparation is unavoidable as the 5 mm thick amorphous glass substrate causes a background that buries the signal of the grains.

**Grain averaged stress and strain.** At the grain level, the slope derived from linear fitting of the normal strain components $\varepsilon_1$ and $\varepsilon_3$ shown in Fig 8b, -0.83, is larger than the biaxial relaxation coefficient of -1.23 reported by Li *et al.* [16]. The forces along the c-axis are set to zero in their model, whereas our stress measurements have shown non-zero elements. The same study by Li *et al.* calculated an increase in the bandgap with the increase of compressive biaxial strain for $\varepsilon_1$ above -1.5% and a decrease for $\varepsilon_1$ below -1.5%. Our $\varepsilon_1$ strain values oscillate between compressive and tensile strain in a range of $[-25; 20] \times 10^{-4}$, which implies that the bandgap is not homogeneous among the grains. The overall bandgap of the film is the result of contributions from all of these grain bandgaps. We also note that the strain is on the order of $10^{-4}$, which agrees with the strains measured in $CuInSe_2$ thin-film solar cells [65].

**Grain properties.** The standard error of the grain variations in angle, length and volume among the measured tetragonal unit cell parameters is 0.008°, $8.8 \times 10^{-4}$ Å, and 0.012 Å$^3$ respectively, demonstrating a high accuracy in the 3DXRD measurements.

The estimation of the grain size based on the intensities of the reflections agrees with the scale shown in the SEM image (Figure 6c). The grain volumes of CZTS grains are larger than those of the ZnS grains.

The main conclusion from the texture analysis is that the <112> poles are preferably aligned to the normal direction of the film, whereas a faint discontinuous ring in the same pole figure could indicate a weak fiber texture. Moreover, the poles <001> and <001> are almost aligned parallel to the surface of the film. This out-of-plane (112) fiber texture has also been observed in a co-sputtered CZTS film [66]. The link between this texture and efficiency is not clearly established, but many studies report the (112) preferred orientation in CZTS films deposited by PLD [67]. In CIGS films, a (200)/(204) preferred orientations yields higher efficiencies than CIGS films with (112) orientations [68].

**Twin boundaries.** The six variants of Σ3-type twin boundaries have also been observed with electron microscopy in CIGS, CGS, and CIS solar cells [59]. Σ3 boundaries have lower defect density compared to random grain boundaries according to electron-beam scattering diffraction (EBSD) and cathodoluminescence (CL) measurements [69].

The formation energies of Σ3 are low, and hence they are common in CZTS films. In our results, 41.2% of the total number of grains are Σ3 twins. The most frequent twin operation is the Σ3{112} that corresponds to the 180° around <221>. Characterization of grain Σ3 {112} twin boundaries in CIGS has been done extensively, revealing a rather benign electronic behavior [69]. Additionally, the formation of Cu vacancies and $In_{Ga}$ antisite defects in Σ3 {112} have been experimentally confirmed [70]. Similarly, one could expect the development of defects in CZTS Σ3{112} twin boundaries. In a theoretical study by Wong et al. [71], Σ3{112} grain boundaries are constructed with $Zn_{Cu}$ and $Sn_{Zn}$ defects based on anion-anion terminations. Such grain boundaries are detrimental and can acts as seeds for secondary phases. According to this model, we could speculate that ZnS grains might be lying close to the Σ3{112} twin boundaries. Moreover, first principle calculations have predicted Cu-poor anion terminated (-1-1-2) surfaces to situate $V_{Cu}$ defects, which are benign for the solar cell performance [72].

## 7.2 3DXRD limitations and new horizons

This paper demonstrates that far-field 3DXRD is suitable for providing comprehensive statistical information about the ensemble of grains in the absorber layer. However, the position of the grain is not resolved. In outlook, one can make mapping of grains in 3D using the 3DXRD scanning modality that employs a smaller X-ray beam (200 nm) and records diffraction patterns at each yz-position of the sample [73]–[75]. Thus, grain positions and strain maps with a higher resolution can be achieved [76]. A drawback with scanning techniques is that the acquisition time increases. However, the next generation of synchrotron sources, such as the Extremely Brilliant Source, EBS, in Grenoble, which was successfully put into operation in Summer 2020, promises an increase in the data acquisition speed of all types of 3DXRD modalities by a factor of 10-50. Moreover, preliminary results on a new full field modality known as High-resolution 3DXRD suggests that 300 nm spatial resolution is within reach [32].

In outlook, combining scanning 3DXRD with X-ray Beam Induced Current (XBIC) and X-ray fluorescence (XRF) could reveal the relation between microstructure and photovoltaics properties of the device and localized elemental composition. By performing *in-operando* studies implementing the mentioned techniques, we could analyse the effects of grain boundaries on PV properties and identify elemental clusters that tend to populate the grain boundaries for passivation. These studies could improve current models for thin-film optimization [71], [77], [78].

## 8 Conclusion

We have characterized the microstructure of a PLD-deposited CZTS absorber layer buried within the stack of layers that constitute a full solar cell device. We demonstrate that 3DXRD can distinguish between phases with nearly identical unit cell parameters. As a result, we found 597 grains; 582 were identified as tetragonal and 15 as cubic. We extracted the strain and stress components both at the sample and at the grain level. We provided extensive statistics of the tetragonal and cubic grains, including the number of grains, sizes, orientations, and twin boundaries of each phase and discussed the relevance of this information for CZTS design.

More generally, the most common photovoltaic thin-film materials are chalcogenides with cubic and tetragonal structures. Structural characterization of these with traditional methods is hampered by the same issues as CZTS. Hence, we propose that the 3DXRD methodology may be applied to index grains of other absorber materials such as CdTe (F-43m) and CIGS (I-42d).

# 9 Acknowledgements

This study was supported by the H2020 European Research Council through the SEEWHI Consolidator grant, ERC-2015-CoG-681881. We acknowledge a travel grant for synchrotron experiments by DANSCATT. We thank Sara Engberg, Andrea Crovetto, Prof. Jørgen Schou and Prof Xiaojing Hao for providing the sample and fruitful discussions. Special thanks are due to Ebtisam Abdellahi for helping with sample preparation and to Henning Osholm Sørensen and Innokenty Kantor for advice on multigrain crystallography. We thank Jonathan Almer, Jun-Sang Park, and Hemant Sharma for their support during the experiment. HFP acknowledge support from the PMP Advanced grant, ERC-2020-Adv-885022 and from the Danish ESS lighthouse on hard materials in 3D, SOLID. This research used resources of the Advanced Photon Source, a U.S. Department of Energy (DOE) Office of Science User Facility, operated for the DOE Office of Science by Argonne National Laboratory under Contract No. DE-AC02-06CH11357.